\documentclass[12pt]{article}
\usepackage{epsfig}
\usepackage{citesort}
\newcommand{\eq}{\begin{equation}}
\newcommand{\eqx}{\end{equation}}
\newcommand{\eqn}{\begin{eqnarray}}
\newcommand{\eqnx}{\end{eqnarray}}
\newcommand{\dt}{\Delta}
\newcommand{\fsn}{$g_1(x,Q^2)\,\,\,\,$}

\newcommand{\ove}{\overline}
\newcommand{\sd}{\displaystyle}
\title{{\bf Low $x$ double $ln^2(1/x)$ resummation effects at the sum rules for
nucleon structure function $g_1$}}
\author{{\sc B. Ziaja}$^{a\,b}$ \footnote{e-mail:beataz@solaris.ifj.edu.pl}\\ \\
$^a$ \it Department of Theoretical Physics,\\
     \it  H.~Niewodnicza\'nski Institute of Nuclear Physics,\\
     \it Radzikowskiego 152, 31-342 Cracow, Poland\\  \\
$^b$ \it High Energy Physics, Uppsala University,\\
        \it P.O. Box 535,S-75121 Uppsala, Sweden }
\date{}
\begin{document}
\maketitle
\noindent
{\bf Abbreviations footnote:}\\ \\
AP-Altarelli-Parisi,\\
LO-leading-order,\\
NLO-next-to-leading order,\\
QCD-quantum chromodynamics,\\
\\ \\
\begin{abstract}
We have estimated the contributions to the moments of polarized nucleon
structure function $g_1(x,Q^2)$ coming from the region of the very low
x ($10^{-5}<x$). Our approach uses the nucleon structure function extrapolated
to the region of low $x$ by the means of the double $ln^2(1/x)$ resummation.
The $Q^2$ evolution of $g_1$ was described by the
unified evolution equations incorporating both the leading order
Altarelli-Parisi evolution at large and moderate $x$, and the double $ln^2(1/x)$ resummation at small $x$.
The moments were obtained by integrating out the extrapolated nucleon structure
function in the region $10^{-5}<x<1$.
\end{abstract}

\section{Introduction}
The sum rules which are expected to be satisfied by the spin-dependent structure function $g_1$ of the nucleon play a
very important role in the theory of the spin-dependent deep inelastic lepton scattering
\cite{JAFFE}.  The sum rules involve (first)  moments of the spin-dependent
structure functions,
and the moment integrals require knowledge of those structure functions in the entire
region $0<x<1$ where, as usual, $x$ denotes the Bjorken variable.  Presently available
experiments cover only the region of large and moderately small values of  $x$
($x>5\cdot10^{-3}$) for reasonably large values of the photon virtuality $Q^2$
($Q^2 > 5 GeV^2$). Hence, a reliable theoretical
estimate of the contributions to the moment integrals coming from the
unmeasured small $x$ region is important for the analysis of the sum rules.

In this paper we propose an extrapolation of the spin-dependent
parton distributions and of the  polarized nucleon structure functions
into the low $x$ region. The extrapolation is based on the double $ln^2(1/x)$
resummation \cite{BZIAJA,BZIAJA1}.  After integrating out the parton distributions
or structure functions, one obtains the low $x$ contribution
to the corresponding moments. The integration interval extends 
from $x\sim10^{-5}$ to $x=1$.
It is assumed here that the small $x$ behaviour of $g_1$ is controlled by the
double $ln^2(1/x)$ resummation. The full analysis of the double $ln^2(1/x)$
resummation effects was performed in detail in Ref.\ \cite{BZIAJA}.
The dominant contribution generating the double logarithmic terms is given
by the ladder diagrams with the quark (antiquark) and gluon exchanges along
the ladder.  The very transparent way of resumming these terms is provided
by the formalism of the unintegrated (spin-dependent) parton
distributions which satisfy the corresponding integral
equations. In \cite{ziaja} we extended this formalism so as to include
the non-ladder  bremsstrahlung terms by adding the suitable higher order
corrections to the kernels of the corresponding integral equations.
We also incorporated the complete leading-order (LO) Altarelli-Parisi
(AP) evolution within this scheme, thus obtaining the unified system of
equations able to analyse simultaneously the parton distributions in the large and
small $x$ regions.
In particular, this formalism allows us to extrapolate dynamically
the spin-dependent structure functions from the region of large and
moderately small values of  $x$,  where they are constrained by
the presently available data to the very small $x$ domain
which can possibly be probed at the polarized HERA \cite{ALBERT}.

This paper is organized as follows. In section 2 we recall briefly
Bjorken and Ellis-Jaffe  sum rules for nucleon structure functions.
In section 3 the unified evolution equations for \fsn \cite{ziaja} 
which embody both
complete LO AP evolution at large values
of $x$ and  the full (ladder and non-ladder) double $ln^2(1/x)$ resummation at small $x$
are discussed in the context of the partonic moment conservation.
It is shown  that in the non-singlet sector the first moments of both baryonic
isovector $g_1^{NS}(x,Q^2)$ and baryonic octet $g_1^{(8)}(x,Q^2)$
are conserved, i.e. they are independent of $Q^2$. It is also shown
that there is no first moment
conservation in the singlet sector. It should be recalled that the first moments
of the non-singlet and octet structure functions acquire their $Q^2$ dependence
only as the result of the next-to-leading order (NLO) quantum chromodynamics
(QCD) effects.  Our formalism extends the LO AP formalism by including
the small $x$ resummation, yet it does not affect the conservation of
the first moments of structure functions.

In Section 4 our predictions for Bjorken and Ellis-Jaffe sum rules obtained
after numerical integration of the respective nucleon
components in the region extending from  very low $x$ ($10^{-5}<x$)
are presented. First moments of nucleon structure functions are calculated and
compared with experimental data \cite{vetterli,pe143,psmc,pemc,ne143,nsmc,ne154,ne142,nhermes}.
In order to estimate the impact of the low $x$ region on the sum rule
integrals and moments, partial contributions from very low $x$ region 
$10^{-5}<x<10^{-3}$ are calculated explicitly.
In our approach we use a simple semi-phenomenological parametrization of
the non-perturbative part of the spin-dependent parton distributions.

In Section 5 the summary of our results is given.

\section{Sum rules for $g_1(x,Q^2)$}

The sum rules for polarized nucleon structure functions are derived from
the space-time representation of scattering amplitudes $T_{ik}^a(x)$
in terms of current commutators \cite{ioffe,ioffe1}~:
\eq
Im\, T_{ik}(x)=\frac{1}{4}\,
\langle p,s \mid \,
\left[j_{i},j_{k}\right]_{antisym.\ }
\,\mid p,s \rangle .
\label{tik0}
\eqx
In the light cone limit $x^2\rightarrow0$, $x_0\rightarrow0$,
which corresponds to parton model kinematics, they reduce to:
\eq
\lim_{x_0\rightarrow 0}\,Im\, T_{ik}(x)=-\epsilon_{ikl}
\langle p,s \mid \frac{1}{3}
\left[j_{5l}^3(0)+\sqrt{\frac{1}{3}}j_{5l}^8(0)\right]
+\frac{2}{9}\,j_{5l}^0(0)
\mid p,s \rangle.
\label{tik}
\eqx
The isospin symmetry determines the proton matrix element of the isovector current
$j_{5l}^3(0)$, and results in the Bjorken sum rule for the non-singlet component
of the nucleon structure functions $g_1^{p,n}(x,Q^2)$ which in LO
approximation reads~:
\eq
\int_{0}^{1}\,g_1^{BJ}(x,Q^2)=g_A/6,
\label{bjor}
\eqx
where $g_1^{BJ}(x,Q^2)\equiv g_1^{p}(x,Q^2)-g_1^{n}(x,Q^2)$,
$g_1^p$, $g_1^n$ are
proton and neutron structure functions respectively, and $g_A\approx 1.257$ is the
neutron $\beta$-decay axial coupling constant. It should be stated clearly
that the Bjorken sum rule acquires also corrections beyond the LO approximation.
Since the formalism of the unified evolution equations we use henceforth
includes only the LO AP evolution, we neglect the NLO correction terms 
both for the Bjorken and the Ellis-Jaffe sum rule.

The Ellis-Jaffe sum rule for baryonic octet follows, if $SU(3)$
flavour symmetry for octet $\beta$-decays is assumed~:
\eq
\int_{0}^{1}\,g_1^{8}(x,Q^2)=(3F-D)/24,
\label{ej1}
\eqx
where~:
\eq
g_1^{8}(x,Q^2)=(\Delta u +\Delta d -2\Delta s)/24,
\label{g8}
\eqx
and $F$, $D$ are octet $\beta$-decay axial coupling constants \cite{ioffe}
fulfilling the relation $(3F-D)/24 \approx 0.0241$. Distributions
$\Delta u$, $\Delta d$, $\Delta s$ denote quark components of the
polarized nucleon. 


\section{Moments of \fsn and double $ln^2(1/x)$ resummation}

Low $x$ behaviour of polarized nucleon structure function is influenced
by double logarithmic  $ln^2(1/x)$ contributions, i.\ e.\  by those terms
of the perturbative expansion, which correspond to the powers of $ln^2(1/x)$
at each order of the expansion \cite{BARTNS,BARTS}.
In what follows we will apply the double $ln^2(1/x)$ resummation scheme
based on the unintegrated parton distributions \cite{BBJK,BZIAJA,MAN}.
Conventional integrated spin-dependent parton  distributions
$\Delta p_l(x,Q^2)$ ($p=q,g$) are related to the unintegrated
parton distributions $f_l(x^{\prime},k^2)$ in the following way ~:
\eq
\Delta p_l(x,Q^2)=\Delta p_l^{(0)}(x)+
\int_{k_0^2}^{W^2}{dk^2\over k^2}\,f_l(x^{\prime}=x(1+{k^2\over Q^2}),k^2),
\label{dpi}
\eqx
where $\Delta p_l^{(0)}(x)$ is the nonperturbative part of the distribution,
$k^2$ denotes the transverse momentum  squared of the probed parton,
$W^2$ is the total energy in the center of mass $W^2=Q^2\,(\frac{1}{x}-1)$,
and index $l$ specifies the parton flavour.
The parameter $k_0^2$ is the infrared cut-off, which will be set equal
to 1 GeV$^2$. The nonperturbative part $\Delta p_l^{(0)}(x)$ can be viewed
upon as originating from the integration over non-perturbative region
$k^2<k_0^2$, i.\ e.\
\eq
\Delta p_l^{(0)}(x)= \int_{0}^{k_0^2}{dk^2\over k^2}f_l(x,k^2).
\label{gint0}
\eqx
The nucleon structure function $g_1(x,Q^2)$ is related in a standard way
to the (integrated) parton distributions describing the parton content of
the polarized nucleon~:
\eqn
g_1^p(x,Q^2)&=&\frac{\langle e^2 \rangle}{2}\,
[g_1^S(x,Q^2)+g_1^{NS,p}(x,Q^2)],\label{g1p}\\
g_1^n(x,Q^2)&=&\frac{\langle e^2 \rangle}{2}\,
[g_1^S(x,Q^2)+g_1^{NS,n}(x,Q^2)],\label{g1n}
\eqnx
where $N_f$ denotes the number of active flavours ($N_f=3$) and 
$\sd \langle e^m\rangle ={1\over N_f}\sum_{l=1}^{N_f}(e_l)^m$. For convenience
we have introduced in (\ref{g1p}), (\ref{g1n}) the non-singlet and singlet
combinations of the spin-dependent quark and antiquark distributions defined
for proton and neutron as:
\eq
g_1^{NS,p(n)}(x,Q^2)= \sum_{l=1}^{N_f} \left({e_l^2\over\langle e^2 \rangle}
- 1\right)(\Delta q^{p(n)}_{l}(x,Q^2) + \Delta \ove q^{p(n)}_{l}(x,Q^2)),
\label{gns}
\eqx
\eq
g_1^S(x,Q^2)= \sum_{l=1}^{N_f}(\Delta q^{\gamma}_{l}(x,Q^2) + \Delta \ove
q^{\gamma}_{l}(x,Q^2)).
\label{gs}
\eqx
In order to consider the Ellis-Jaffe sum rule we shall also analize
the baryon octet structure function $g^8(x,Q^2)$, defined by equation (\ref{g8}).

\begin{figure}[t]
    \centerline{
     \epsfig{figure=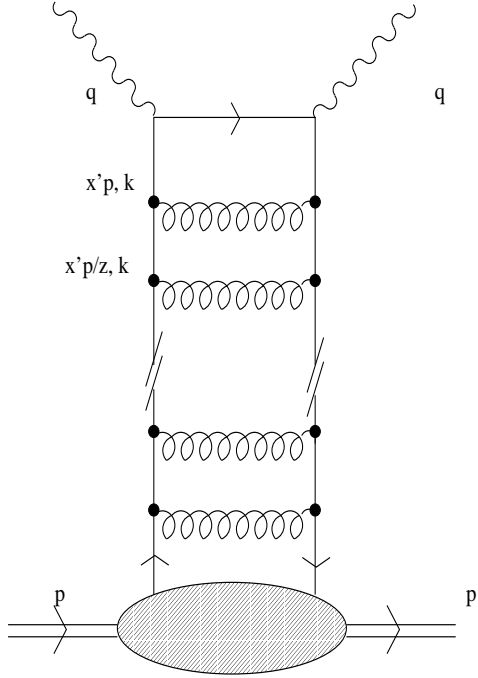,height=9cm,width=6cm}
               }
     \caption{Ladder diagram generating the
     double logarithmic terms in the non-singlet component of the 
     spin structure function.}
\label{f1}
\end{figure}
%
\begin{figure}[t]
    \centerline{
     \epsfig{figure=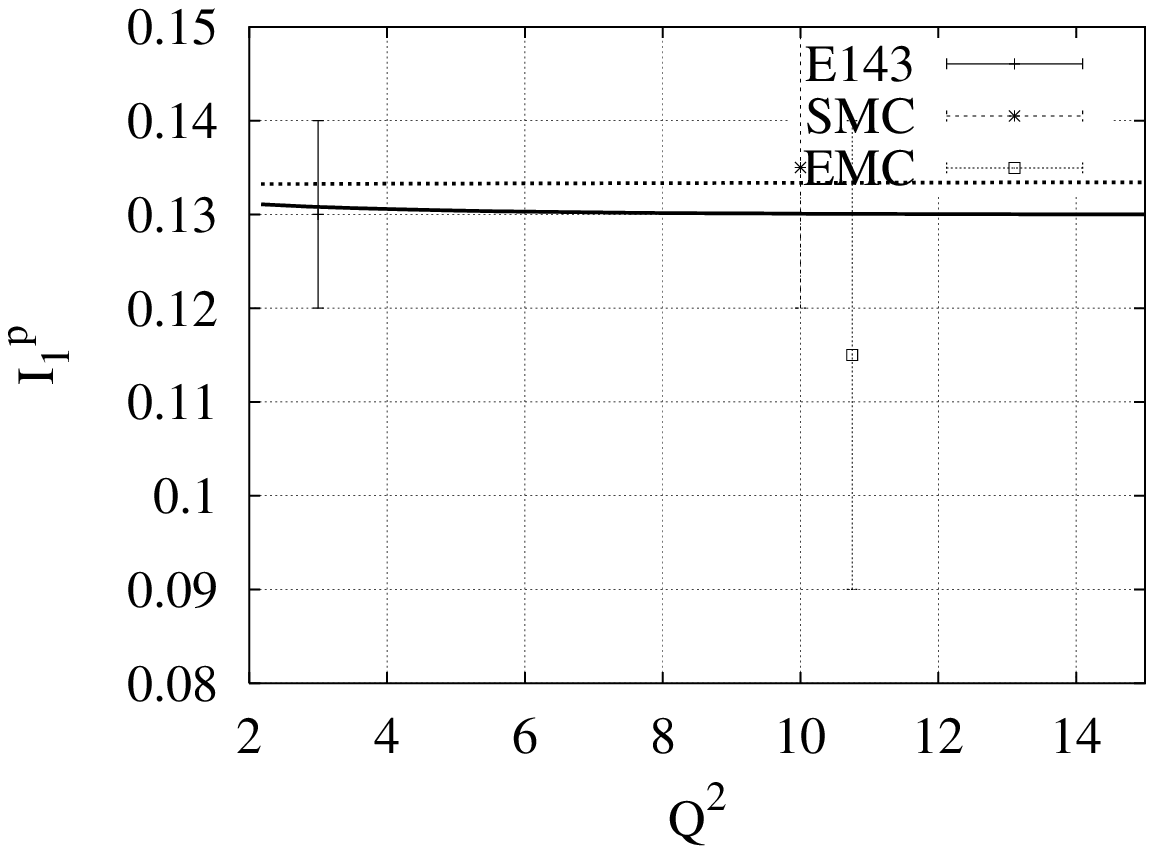,width=10cm}
               }
     \caption{The moment $I_1^p$ of proton structure function $g_1^p(x,Q^2)$
     plotted as a function of $Q^2$. Solid line denotes results obtained
     from the unified evolution including the full double logarithmic
     resummation $ln^2(1/x)$, dashed line shows pure AP evolution, 
     dotted line corresponds to the non-perturbative
     input (overlaps with AP results). Experimental data are denoted: 
     E143 \cite{pe143} with vertical bars,
     SMC \cite{psmc} with stars, EMC \cite{pemc} with white squares.}
\label{f2}
\end{figure}
%
%
\begin{figure}[t]
    \centerline{
     \epsfig{figure=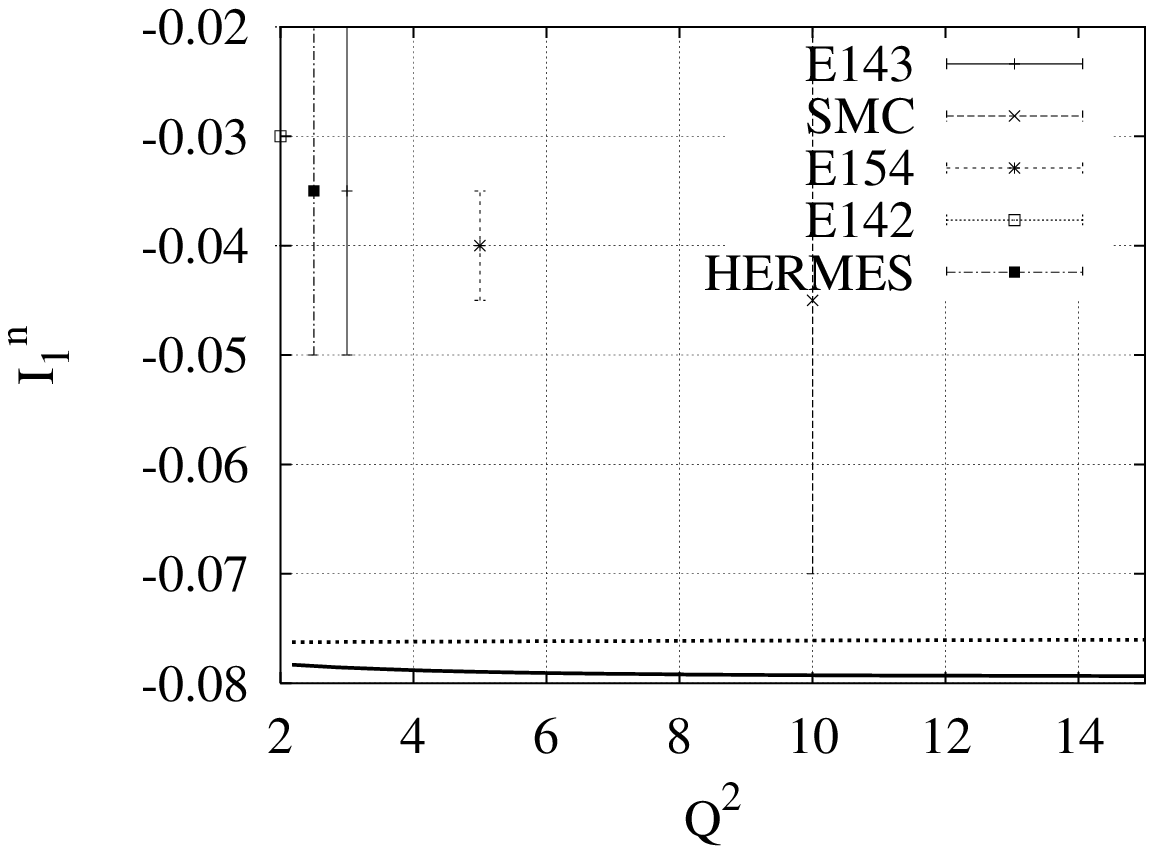,width=10cm}
               }
     \caption{The moment $I_1^n$ of neutron structure function $g_1^n(x,Q^2)$
     plotted as a function of $Q^2$. Solid line denotes results obtained
     from the unified evolution including the full double logarithmic
     resummation $ln^2(1/x)$, dashed line
     shows pure AP evolution, dotted line corresponds to the non-perturbative
     input (overlaps with AP results). Experimental data are denoted: 
     E143 \cite{ne143} with vertical bars, SMC \cite{nsmc} with crosses, 
     E154 \cite{pemc} with stars, E142
     \cite{ne142} with white squares, HERMES \cite{nhermes} with black
     squares.}
\label{f3}
\end{figure}
%
\begin{figure}[t]
    \centerline{
     \epsfig{figure=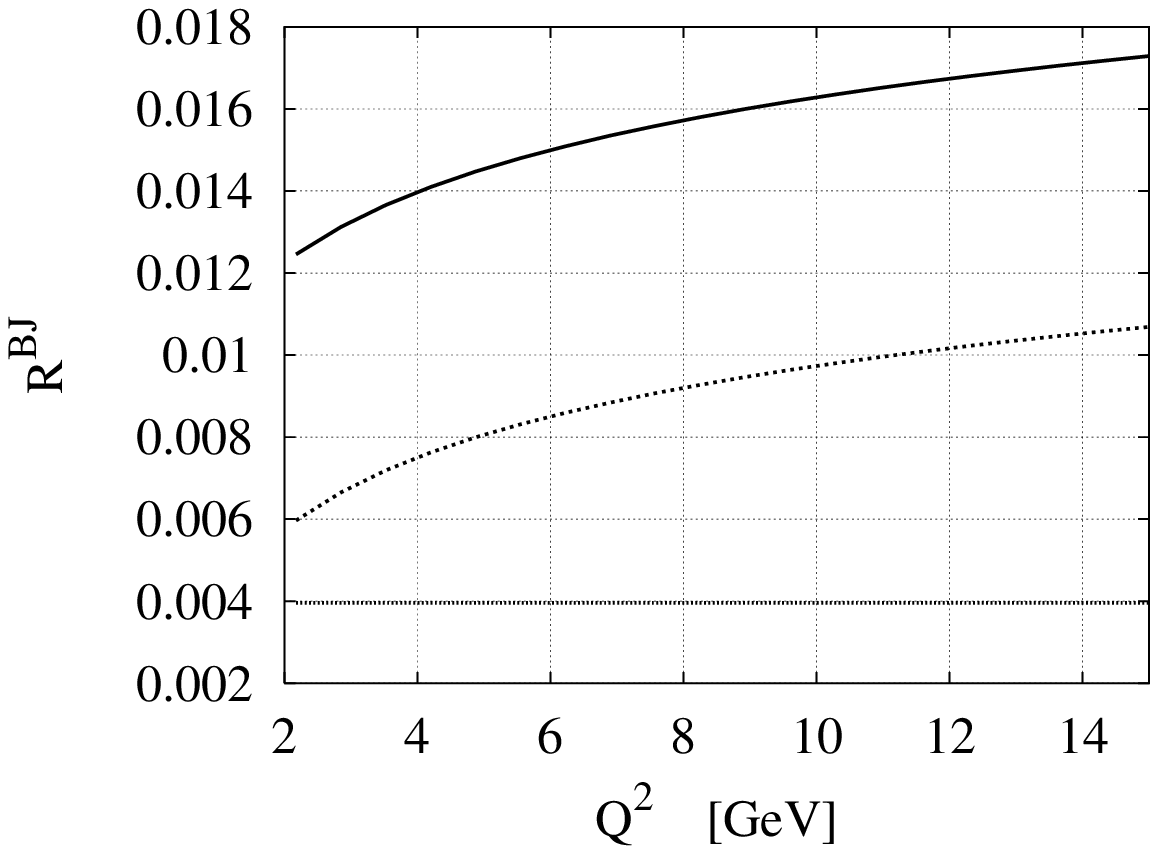,width=11cm}
               }
     \caption{The ratio $R^{BJ}(Q^2)$ for Bjorken sum rule integral plotted
     as a function of $Q^2$. Solid line denotes results obtained
     from the unified evolution including the full double logarithmic
     resummation $ln^2(1/x)$, dashed line
     shows pure AP evolution, dotted line corresponds to the non-perturbative
     input.}
\label{f4}
\end{figure}
%
%
\begin{figure}[t]
    \centerline{
     \epsfig{figure=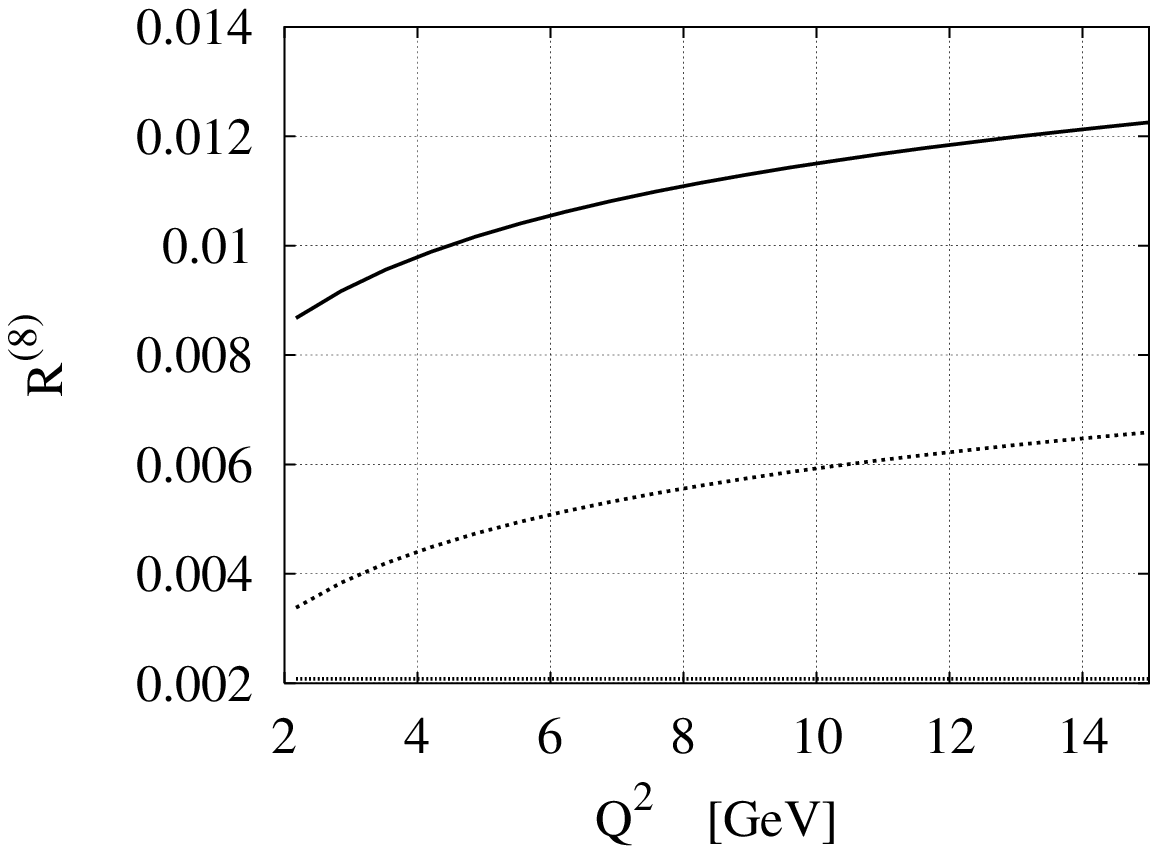,width=11cm}
               }
     \caption{The ratio $R^{8}(Q^2)$ for Ellis-Jaffe sum rule integral plotted
     as a function of $Q^2$. Solid line denotes results obtained
     from the unified evolution including the full double logarithmic
     resummation $ln^2(1/x)$, dashed line
     shows pure AP evolution, dotted line corresponds to the non-perturbative
     input.}
\label{f5}
\end{figure}
%
\begin{figure}[t]
    \centerline{
     \epsfig{figure=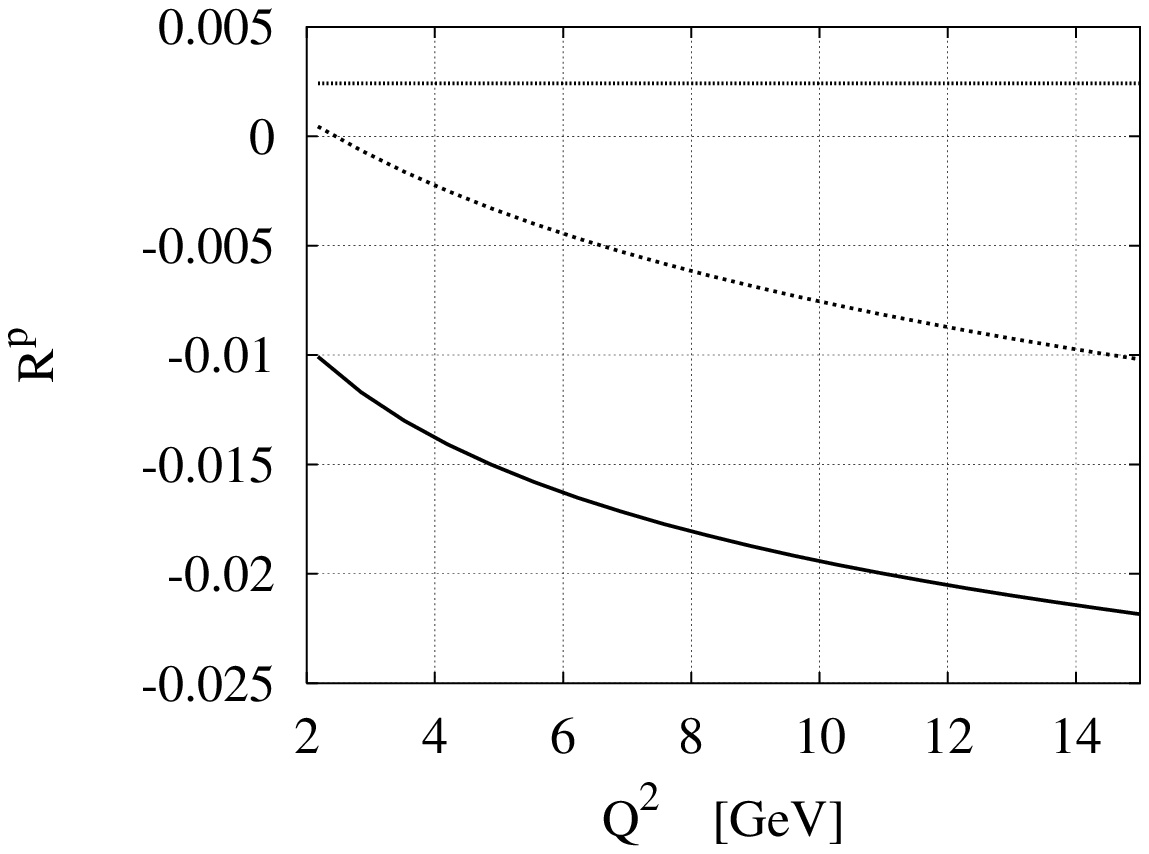,width=11cm}
               }
     \caption{The ratio $R^{p}(Q^2)$ for the first moment of proton structure
     function plotted
     as a function of $Q^2$. Solid line denotes results obtained
     from the unified evolution including the full double logarithmic
     resummation $ln^2(1/x)$, dashed line
     shows pure AP evolution, dotted line corresponds to the non-perturbative
     input.}
\label{f6}
\end{figure}
%
\begin{figure}[t]
    \centerline{
     \epsfig{figure=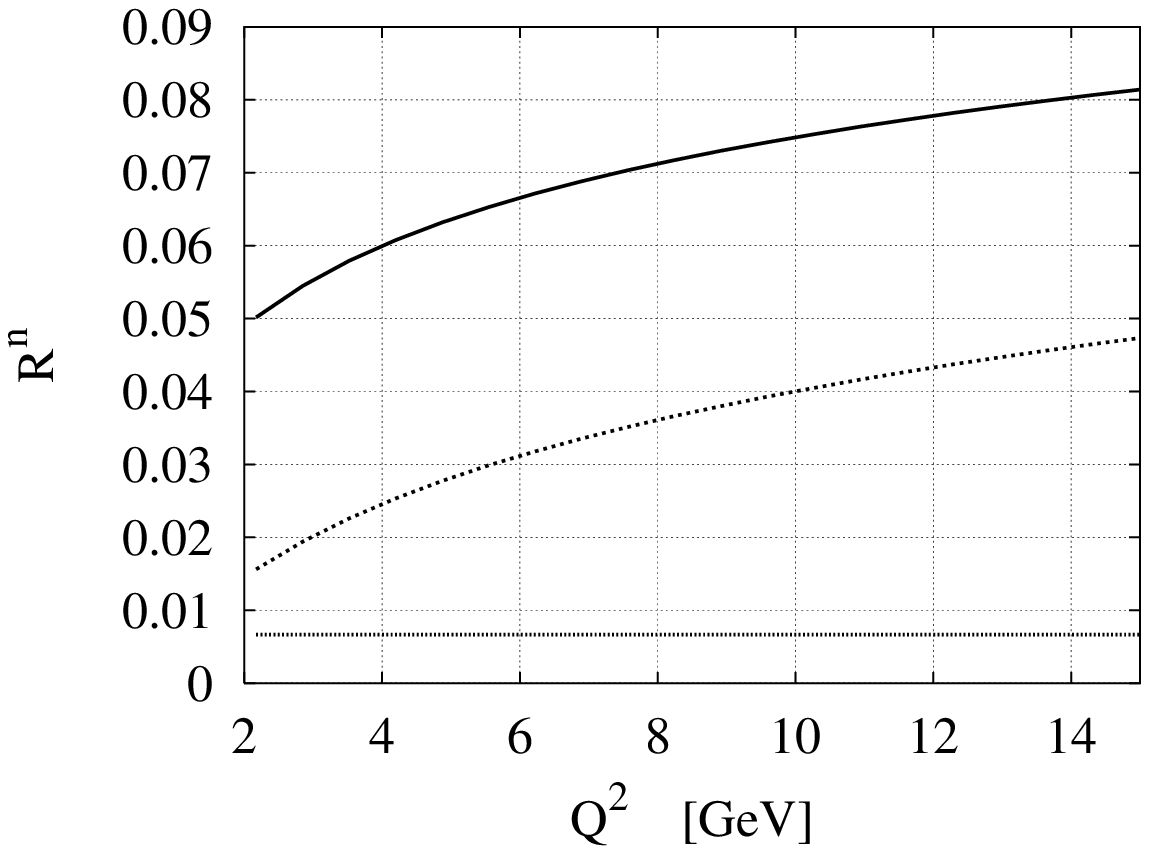,width=11cm}
               }
     \caption{The ratio $R^{n}(Q^2)$ for the first moment of neutron structure
     function plotted
     as a function of $Q^2$. Solid line denotes results obtained
     from the unified evolution including the full double logarithmic
     resummation $ln^2(1/x)$, dashed line
     shows pure AP evolution, dotted line corresponds to the non-perturbative
     input.}
\label{f7}
\end{figure}

The full contribution to the double $ln^2(1/x)$ resummation comes from
the ladder diagrams with quark and gluon exchanges along the ladder
(cf.\ Fig.\ \ref{f1}) and the non-ladder bremsstrahlung diagrams \cite{QCD,QCD1}.
The latter ones are obtained from the ladder diagrams by adding to them
soft bremsstrahlung gluons or soft quarks \cite{BARTNS,BARTS,QCD,QCD1}.
They generate the infrared corrections to the ladder contribution.

The relevant region of phase space generating the double
$ln^2(1/x)$ resummation from  ladder diagrams corresponds
to ordered $k_n^2/x_n$, where $k_n^2$ and  $x_n$ denote respectively the
transverse momenta squared and longitudinal momentum fractions of the
proton, carried by  partons  exchanged along the ladder \cite{QED,QED1}. It is
in contrast to the LO AP  evolution alone, which corresponds to ordered
transverse momenta.

The structure of the corresponding integral equations describing unintegrated
distributions $f_{NS}(x{^\prime},k^2), f_{S}(x{^\prime},k^2)$ and
$f_g(x{^\prime},k^2)$ for ladder diagram contribution read \cite{BZIAJA}:
\eq
f_{NS}(x^{\prime},k^2)=f^{(0)}_{NS}(x^{\prime},k^2) +
{\alpha_S \over 2 \pi} \Delta P_{qq}(0)
\int_{x^{\prime}}^1 {dz\over z}
\int_{k_0^2}^{k^2/z}
{dk^{\prime 2}\over k^{\prime 2}}
f_{NS}\left({x^{\prime}\over z},k^{\prime 2}\right),
\label{dlxns}
\eqx
{\footnotesize
\eqn
f_{S}(x^{\prime},k^2)=f^{(0)}_{S}(x^{\prime},k^2)+
{\alpha_S\over 2 \pi}
\int_{x^{\prime}}^1 {dz\over z}
\int_{k_0^2}^{k^2/z}
{dk^{\prime 2}\over k^{\prime 2}}
\left[\Delta P_{qq}(0)
f_{S}\left({x^{\prime}\over z},k^{\prime 2}\right)+
\Delta P_{qg}(0)
f_{g}\left({x^{\prime}\over z},k^{\prime 2}\right)\right]\nonumber,
\eqnx}
{\footnotesize
\eqn
f_{g}(x^{\prime},k^2)=f^{(0)}_{g}(x^{\prime},k^2) +
{\alpha_S\over 2 \pi}
\int_{x^{\prime}}^1 {dz\over z}
\int_{k_0^2}^{k^2/z}
{dk^{\prime 2}\over k^{\prime 2}}
\left[\Delta P_{gq}(0)
f_{S}\left({x^{\prime}\over z},k^{\prime 2}\right)+
\Delta P_{gg}(0)
f_{g}\left({x^{\prime}\over z},k^{\prime 2}\right)\right]\nonumber\\
\label{dlxsg}
\eqnx}
with splitting functions $\Delta P_{ij}(0) \equiv \Delta P_{ij}(z=0)$
equal to~:
\begin{eqnarray}
{\bf \dt P(0)} \equiv
\left( \begin{array}{cc} \Delta P_{qq}(0) & \Delta P_{qg}(0)\\
\Delta P_{gq}(0) & \Delta P_{gg}(0) \\ \end{array} \right ) =
\left( \begin{array}{cc}  {{N_C^2-1 \over 2N_C}} & - N_F \\
                          {{N_C^2-1 \over  N_C}} &  4N_C  \\ \end{array}\right),
\label{dpij}
\end{eqnarray}
\noindent
where $\alpha_S$ denotes the QCD coupling, which at the moment is treated as a fixed 
parameter.
The variables $k^2$($k^{\prime 2}$) denote the transverse momenta
squared of the quarks (gluons), exchanged along the ladder.
For the parton distributions in a hadron the inhomogeneous driving terms
$f^{(0)}_l(x',k^2)$ are entirely determined by the non-perturbative
parts $\Delta p_i^{(0)}(x')$ of the spin-dependent parton distributions.

Besides the ladder diagrams contributions, the double logarithmic resummation
does also acquire corrections from the non-ladder bremsstrahlung
contributions.  It has been shown in Ref. \cite{BZIAJA} that these
contributions can be included
by adding the higher order terms to the kernels of  integral equations
(\ref{dlxns}), (\ref{dlxsg}).
These terms can be obtained from the matrix:
$ \sd \Biggl[ {\tilde  {\bf F}_8 }/\omega^2 \Biggr](z) \bf G_0$,
where $\sd \Biggl[ \frac{\tilde  {\bf F}_8 }{\omega^2} \Biggr](z)$
denote the inverse Mellin transform of the octet partial wave matrix
(divided by $\omega^2$ ), and the matrix  $\bf G_0$ reads:
\eqn
{\bf G}_0 &=&\left( \begin{array}{cc}  {N_c^2-1 \over 2N_c} & 0  \\
                                                    0 & N_c  \\ \end{array}
						    \right).	
\label{g0}						
\eqnx
Following Ref. \cite{BZIAJA}, we shall use the Born approximation for the octet
matrix~:
\eq
\Biggl[\frac{\tilde {\bf F}_8^{Born}}{\omega^2}\Biggr](z)=
4\pi^2 \ove \alpha_S{\bf M}_8 ln^2 (z),
\label{born}
\eqx
where $\bf M_8$ is the splitting functions matrix in colour octet
t-channel~:
\eqn
{\bf M}_8 &=&\left( \begin{array}{cc} -{1 \over 2N_c} & -{N_F \over 2}\\
                                                N_c & 2N_c \\ \end{array} \right ).
\label{m8g0}
\eqnx

In the region of large values of $x$ the integral equations
(\ref{dlxns}), (\ref{dlxsg}) describing pure double logarithmic resummation
$\ln^2(1/x)$, even completed by including non-ladder contributions,
are inaccurate. In this region one should use the conventional AP  equations
\cite{STRATMANN1,ALTAR,AP} with the complete splitting functions $\Delta P_{ij}(z)$
and not restrict oneself to the effect generated by their $z\rightarrow 0$ part.
Following Refs.\ \cite{BBJK,BZIAJA}, we do therefore extend equations
(\ref{dlxns}), (\ref{dlxsg}), and add to their right-hand-side(s) the
contributions coming from the remaining parts of the splitting functions
$\Delta P_{ij}(z)$. We also allow coupling $\alpha_S$ to run, setting $k^2$
as the relevant scale. In this way we obtain unified system of equations,
which contain both the complete leading order AP  evolution and the double
logarithmic $ln^2(1/x)$ effects at low $x$. These equations
are listed in Appendix A.


\subsection{Conservation of moments}

In order to get information about the moments of spin structure functions,
we will follow the technique proposed in \cite{BBJK}.
First we integrate the integrated parton distributions (\ref{dpi}) over $x$~:
\eq
\int_0^1\,dx\,\Delta p_l(x,Q^2)=
\int_0^1\,dx\,\Delta p_l^{(0)}(x)+
\int_{k_0^2}^{\infty}{dk^2\over {k^2(1+\frac{k^2}{Q^2})} }
\,\int_0^1\,dx\,f_l(x,k^2).
\label{dpimom}
\eqx
Let us denote
${\ove \Delta p_l(Q^2)}\,\equiv\,\int_0^1\,dx\,\Delta p_l(x,Q^2)$, and
respectively ${\ove f_l(k^2)}\,\equiv\,\int_0^1\,dx\,f_l(x,k^2)$. Since
structure functions $g_1^{NS\,(S)}(x,Q^2)$ (\ref{gns}), (\ref{gs}) are linear
combinations of (\ref{dpi}), their moments may be obtained as~:
\eqn
{\ove g_1^{NS}(Q^2)}&=&{\ove g_1^{NS,(0)}}+
\int_{k_0^2}^{\infty}{dk^2\over {k^2(1+\frac{k^2}{Q^2})} }
{\ove f_{NS}(k^2)},\label{nsmom}\\
{\ove g_1^{S}(Q^2)}&=&{\ove g_1^{S,(0)}}+
\int_{k_0^2}^{\infty}{dk^2\over {k^2(1+\frac{k^2}{Q^2})} }
{\ove f_{S}(k^2)}\label{smom},
\eqnx
where ($i=NS,S$)~:
\eqn
{\ove g_1^{i}}    &=&\int_0^1\,dx\,g_1^{i}(x),\\
{\ove g_1^{i,(0)}}&=&\int_0^1\,dx\,g_1^{i,(0)}(x).
\eqnx

Furthermore, for non-singlet sector it was proven \cite{BBJK} (see
Appendix B) that the moments of $f^{(0)}_{NS}(x^{\prime},k^2)$ (\ref{fns0})
vanish independently of the input $g_1^{NS(0)}(x)$~:
\eq
{\ove f_{NS}^{(0)}(k^2)}=0.
\label{nsmom00}
\eqx
Therefore the unified equation for moments ${\ove f_{NS}(k^2)}$, obtained from
equation (\ref{unifns}) after integration over $x$, reduces to the
integral equation with inhomogeneous term equal to $0$. Its solution then
reads~:
\eq
{\ove f_{NS}(k^2)}=0.
\label{nsmom0}
\eqx

For the singlet sector the situation gets more complicated. Although the quark
singlet moment~:
\eq
{\ove f_{S}^{(0)}(k^2)}=0
\label{fsmom0}
\eqx
vanishes again (see Appendix C), the moment of the input gluon distribution
${\ove f_{g}^{(0)}(k^2)}$ takes a non-zero value which reads~:
\eq
{\ove f_{g}^{(0)}(k^2)}=\frac{\alpha_S(k^2)}{2\pi}\,
\left[2\,{\ove g_1^{S,(0)} }+({11\over 2}-{N_F\over 3}){\ove \Delta p_g^{(0)}}\right]
\label{fgmom0}.
\eqx
Hence, the unified equations for quark singlet and gluon moments, obtained
from coupled  equations  (\ref{unifsea}), (\ref{unifglue}) get a non-zero
inhomogeneous term in the gluon sector, and the moment
${\ove f_{S}(k^2)}$ is non-vanishing as well.

The properties of partonic moments resulting from the unified equations (\ref{unifns}),
(\ref{unifsea}), (\ref{unifglue}) after transforming them into the moment
space have clear implications for Bjorken and Ellis-Jaffe sum rules.
Both Bjorken and Ellis-Jaffe sum rules (\ref{bjor}), (\ref{ej1}) concern
evolution in the non-singlet sector. Hence, if one assumes input distributions
$g_1^{BJ,(0)}$ and $g_1^{8,(0)}$ fulfilling the requirements (\ref{bjor}),
(\ref{ej1}), it follows from relations (\ref{nsmom}),
{(\ref{nsmom0}) that~:
\eqn
{\ove g_1^{BJ}(Q^2)}&=&{\ove g_1^{BJ,(0)}}=const\label{bjoru},\\
{\ove g_1^{8}(Q^2)} &=&{\ove g_1^{8, (0)}}=const\label{ej1u}
\eqnx
are conserved throughout whole $Q^2$ evolution.

The non-vanishing unintegrated gluon input (\ref{fgmom0}) implies that
there is no explicit conservation of ${\ove g_1^{S}(Q^2)}$
( and ${\bar \Delta g(Q^2)}$ ) during $Q^2$ evolution.
However, the conservation may be achieved implicitly
by imposing the negative input gluon distribution ${\ove \Delta p_g^{(0)}}$ to 
fulfill the requirement~:
\eq
2\,{\ove g_1^{S,(0)} }+({11\over 2}-{N_F\over 3}){\ove \Delta p_g^{(0)}}=0.
\label{gluon0}
\eqx
This is not a physical case but this shows that the moments are very
sensitive to the gluon input which, in fact, has not an established
phenomenological parametrization because of the lack of experimental
data. Therefore, there is still possible to influence the $Q^2$ evolution of  
${\ove g_1^{S,(0)} }$ and henceforth, the evolution of $\Gamma^{1,p,n}(Q^2)$
by manipulating the input gluon distribution.

\section{Numerical results for sum rules}

We solved the unified equations (\ref{unifns}), (\ref{unifsea}),
(\ref{unifglue}),
assuming the following simple parametrization of the
input distributions:
\begin{equation}
\Delta p_i^{(0)}(x)=N_i (1-x)^{\eta_i}
\label{dpi0}
\end{equation}
with $\eta_{u_v}=\eta_{d_v}=3,$  $\eta_{\ove u} = \eta_{\ove s} = 7 $
and $\eta_g=5$.  The normalisation constants $N_i$ were determined
by imposing the Bjorken sum rule for $\Delta u_v^{(0)}-\Delta d_v^{(0)}$
and requiring that the first moments of
all other distributions are the same as those determined from  the recent
QCD analysis \cite{STRATMAN}. All distributions $\Delta p_i^{(0)}(x)$
behave as $x^0$ in the limit $x\rightarrow 0$ that corresponds to the implicit
assumption that the Regge poles which
should control the small $x$ behaviour of $g_1^{(0)}$ have their intercept equal
to $0$.

It was checked that the parametrization (\ref{dpi0}) combined with
equations (\ref{dpi}), (\ref{gns}), (\ref{gs}), (\ref{unifns}),
(\ref{unifsea}), (\ref{unifglue}) gives reasonable description of the recent
SMC data on $g_1^{BJ}(x,Q^2)$ and on $g_1^p(x,Q^2)$ \cite{nsmc}.
After integrating out the respective parton distributions, we found
that the moments ${\ove g_1^{BJ}(Q^2)}$ and ${\ove g_1^{8}(Q^2)}$
are conserved during $Q^2$ evolution  with a good accuracy (not shown). 
The Bjorken sum rule is conserved explicitly, due to the choice of the input
(${\ove g_1^{BJ,(0)}(Q^2)}\approx 0.2095$). For the  ${\ove g_1^{8}(Q^2)}$
moment, the input was chosen to give  ${\ove g_1^{8,(0)}(Q^2)} \approx 0.016$,
which was implicitly in disagreement with hyperon $\beta$-decay data (\ref{g8}).

We also investigated the $Q^2$ evolution of  the first moments of
$g_1^{p,n}(x,Q^2)$ and compared them with experimental data (see Figs.\
\ref{f2}, \ref{f3}). First moments of $g_1^{p}(x,Q^2)$ agree well with
the available experimental data both for moments obtained after performing
LO AP evolution of non-perturbative input and for moments
obtained after solving the unified evolution equation (cf.\ Fig.\ \ref{f2}).
On the contrary, our predictions for the first moments of polarized neutron
structure function $g_1^{n}(x,Q^2)$ obtained from both the unified and the
AP evolution are below the experimental data (cf.\ Fig.\ \ref{f3}).
The discrepancy may be due to the fact that we consider only the LO AP
evolution of the partonic moments. Also the Bjorken sum rule is considered
at LO accuracy. Since the AP evolution dominates in the
region of  moderate and large $x$, applying it with the leading order
(parton model) accuracy may be not sufficient to reproduce the experimental
data.

Moreover, we estimated the magnitude of contribution from the very low $x$
region ($10^{-5}<x<10^{-3}$) to the moments of polarized nucleon
structure function. It was achieved by comparing the partial contributions
with the total moments obtained after integrating $g_1^{p,n}$ over $x$
extending from $10^{-5}$ to $1$. The calculated ratios,
$\sd R^{i} \equiv {\ove g_1^{i,L}(Q^2)}/{\ove g_1^{i}(Q^2)}$ ($i=BJ,8,p,n$),
where
$\sd {\ove g_1^{i,L}(Q^2)}\,\equiv\,\int_{10^{-5}}^{10^{-3}}\,dx\,g_1^{i}
(x,Q^2)$, are plotted in Figs.\ \ref{f4}, \ref{f5}, \ref{f6}, \ref{f7}.
For Bjorken integral (BJ) (\ref{bjor}) the maximal $\mid R^{BJ}\mid $ 
is $\sim0.02$, and for baryonic octet ($8$), $\mid R^{8}\mid \sim0.01$.
For proton (p) the maximal contribution of low $x$ region to the first
moment of the proton structure function is $\mid R^p \mid \sim0.02$, and
for neutron (n), $\mid R^n \mid \sim 0.08$.

\section{Summary}

To sum up, we have estimated the contributions to the moments of polarized
nucleon structure function $g_1(x,Q^2)$ coming from the region of the very low
x ($10^{-5}<x$). Our approach used the nucleon structure function extrapolated
to the region of low $x$ by the means of the double $ln^2(1/x)$ resummation
which dominates in this region. The $Q^2$ evolution of $g_1$ was described by
the unified evolution equations, which incorporated both the LO AP evolution
manifesting at large and moderate $x$, and double $ln^2(1/x)$ resummation
dominating at small $x$.
These moments were obtained by integrating out the extrapolated nucleon structure
function in the region $10^{-5}<x<1$.

Moments of proton structure function estimated for $2<Q^2<15$ GeV$^2$ were
found in agreement with experimental data. The explicitly evaluated contribution
of low $x$ region obtained via integrating out the $g_1(x,Q^2)$ in the
region of very low $x$ ($10^{-5}<x<10^{-3}$) was found to constitute about 2\%
of the total proton moment for $2<Q^2<15$ GeV$^2$.

Moments of neutron structure function estimated for $2<Q^2<15$ GeV$^2$ were
found to lie below the experimental data. This may be due to the fact that
the unified evolution equations contain only leading order AP kernels.
The contribution of the  very low $x$ region
($10^{-5}<x<10^{-3}$) was found to constitute around 8\% of the total
neutron moment for $2<Q^2<15$ GeV$^2$.

Contributions from very low $x$ region to Bjorken and Ellis-Jaffe sum rules
were of the order $1$ \% and $2$ \% respectively.

This implies that the contribution of low $x$ region enters the moments of
polarized nucleon structure function at the level of 10 \% at most. The
contribution increases with increasing $Q^2$. We expect that the improvements 
of the model needed to describe accurately the neutron data will possibly 
affect only the normalization of the corresponding integrals, and they will 
not change significantly the estimate of the relative contributions coming 
from the low x region.

\section*{Appendix A}
The corresponding system of equations reads~:
{\small
\eqn
f_{NS}(x^{\prime},k^2)=f^{(0)}_{NS}(x^{\prime},k^2)
&+&
{\alpha_S(k^2)\over 2 \pi}{4\over 3}
\int_{x^{\prime}}^1 {dz\over z}
\int_{k_0^2}^{k^2/z}
{dk^{\prime 2}\over k^{\prime 2}}
f_{NS}\left({x^{\prime}\over z},k^{\prime 2}\right)\label{unifns}\\
&&\hspace*{10ex}{\bf (\hspace*{3ex}Ladder\hspace*{3ex})}\nonumber\\
&+&{\alpha_S(k^2)\over 2\pi}\int_{k_0^2}^{k^2}{dk^{\prime 2}\over
k^{\prime 2}}{4\over 3}\int _{x^{\prime}}^1
{dz\over z} {(z+z^2)f_{NS}({x^{\prime}\over z},k^{\prime 2})-
2zf_{NS}(x^{\prime},k^{\prime 2})\over 1-z}\nonumber\\
&+&{\alpha_S(k^2)\over 2\pi}\int_{k_0^2}^{k^2}{dk^{\prime 2}\over
k^{\prime 2}}\left[2 +
{8\over 3} ln(1-x^{\prime})\right]f_{NS}(x^{\prime},k^{\prime 2})\nonumber\\
&&\hspace*{10ex}{\bf(\hspace*{3ex}Altarelli - Parisi \hspace*{4ex})}\nonumber\\
&-&{\alpha_S(k^2)\over 2\pi}
\int_{x^{\prime}}^1 {dz\over z}
\Biggl(
\Biggl[ \frac{\tilde  {\bf F}_8 }{\omega^2} \Biggr](z)
\frac{ {\bf G}_0 }{2\pi^2}
\Biggr)_{qq}
\int_{k_0^2}^{k^2}
{dk^{\prime 2}\over k^{\prime 2}}
f_{NS}\left({x^{\prime}\over z},k^{\prime 2}\right)\nonumber\\
&-&{\alpha_S(k^2)\over 2\pi}
\int_{x^{\prime}}^1 {dz\over z}
\int_{k^2}^{k^2/z}
{dk^{\prime 2}\over k^{\prime 2}}
\Biggl(
\Biggl[\frac{\tilde   {\bf F}_8 }{\omega^2} \Biggr]
\Biggl(\frac{k^{\prime 2}}{k^2}z \Biggr)\frac{ {\bf G}_0 }{2\pi^2}
\Biggr)_{qq}
f_{NS}\left({x^{\prime}\over z},k^{\prime 2}\right),\nonumber\\
&&\hspace*{10ex}{\bf(Non-ladder)}\nonumber
\eqnx
}
{\small
\eqn
f_{S}(x^{\prime},k^2)=f^{(0)}_{S}(x^{\prime},k^2)
&+&
{\alpha_S(k^2)\over 2 \pi}
\int_{x^{\prime}}^1 {dz\over z}
\int_{k_0^2}^{k^2/z}
{dk^{\prime 2}\over k^{\prime 2}}
{4\over 3}
f_{S}\left({x^{\prime}\over z},k^{\prime 2}\right)\nonumber\\
&-&{\alpha_S(k^2)\over 2 \pi}\int_{x^{\prime}}^1 {dz\over z}
\int_{k_0^2}^{k^2/z}
N_F{dk^{\prime 2}\over k^{\prime 2}}f_{g}
\left({x^{\prime}\over z},k^{\prime 2}\right)\nonumber\\
&&\hspace*{10ex}{\bf (\hspace*{3ex}Ladder\hspace*{3ex})}\nonumber\\
&+&{\alpha_S(k^2)\over 2 \pi}\int_{k_0^2}^{k^2}{dk^{\prime 2}\over
k^{\prime 2}}{4\over 3}\int _{x^{\prime}}^1
{dz\over z} {(z+z^2)f_{S}({x^{\prime}\over z},k^{\prime 2})-
2zf_{S}(x^{\prime},k^{\prime 2})\over 1-z}\nonumber\\
&+&{\alpha_S(k^2)\over 2 \pi}\int_{k_0^2}^{k^2}{dk^{\prime 2}\over
k^{\prime 2}}\left[ 2 +
{8\over 3} ln(1-x^{\prime})\right]
f_{S}(x^{\prime},k^{\prime 2}) \nonumber\\
&+&{\alpha_S(k^2)\over 2 \pi} \int_{k_0^2}^{k^2}{dk^{\prime 2}\over
k^{\prime 2}}\int _{x^{\prime}}^1
{dz\over z} 2z N_F f_g({x^{\prime}\over z},k^{\prime 2})\nonumber\\
&&\hspace*{10ex}{\bf(\hspace*{3ex}Altarelli - Parisi \hspace*{4ex})}\nonumber\\
&-&{\alpha_S(k^2)\over 2\pi}
\int_{x^{\prime}}^1 {dz\over z}
\Biggl(
\Biggl[ \frac{\tilde  {\bf F}_8 }{\omega^2} \Biggr](z)
\frac{ {\bf G}_0 }{2\pi^2}
\Biggr)_{qq}
\int_{k_0^2}^{k^2}
{dk^{\prime 2}\over k^{\prime 2}}
f_{S}\left({x^{\prime}\over z},k^{\prime 2}\right)\label{unifsea}\\
&-&{\alpha_S(k^2)\over 2\pi}
\int_{x^{\prime}}^1 {dz\over z}
\int_{k^2}^{k^2/z}
{dk^{\prime 2}\over k^{\prime 2}}
\Biggl(
\Biggl[\frac{\tilde   {\bf F}_8 }{\omega^2} \Biggr]
\Biggl(\frac{k^{\prime 2}}{k^2}z \Biggr)\frac{ {\bf G}_0 }{2\pi^2}
\Biggr)_{qg}
f_{g}\left({x^{\prime}\over z},k^{\prime 2}\right),\nonumber\\
&&\hspace*{10ex}{\bf(Non-ladder)}
\nonumber
\eqnx
}
{\small
\eqn
f_{g}(x^{\prime},k^2)=f^{(0)}_{g}(x^{\prime},k^2) &+&
 {\alpha_S(k^2)\over 2 \pi}
\int_{x^{\prime}}^1 {dz\over z}
\int_{k_0^2}^{k^2/z}
{dk^{\prime 2}\over k^{\prime 2}}
{8\over 3}
f_{S}\left({x^{\prime}\over z},k^{\prime 2}\right)\nonumber\\
&+&{\alpha_S(k^2)\over 2 \pi}
\int_{x^{\prime}}^1 {dz\over z}
\int_{k_0^2}^{k^2/z}
{dk^{\prime 2}\over k^{\prime 2}}
12 f_{g}\left({x^{\prime}\over z},k^{\prime 2}\right)\nonumber\\
&&\hspace*{10ex}{\bf (\hspace*{3ex}Ladder\hspace*{3ex})}\nonumber\\
&+&{\alpha_S(k^2)\over 2 \pi}
\int_{k_0^2}^{k^2}
{dk^{\prime 2}\over k^{\prime 2}}\int_{x^{\prime}}^1 {dz\over z}
(-{4\over 3})zf_{S}
\left({x^{\prime}\over z},k^{\prime 2}\right)\nonumber\\
&+&{\alpha_S(k^2)\over 2 \pi}
\int_{k_0^2}^{k^2}
{dk^{\prime 2}\over k^{\prime 2}} \int_{x^{\prime}}^1 {dz\over z}
6z\left[{f_{g}
\left({x^{\prime}\over z},k^{\prime 2}\right)- f_{g}
(x^{\prime},k^{\prime 2})\over 1-z} -2f_{g}
\left({x^{\prime}\over z},k^{\prime 2}\right)\right]\nonumber\\
&+&{\alpha_S(k^2)\over 2 \pi}
\int_{k_0^2}^{k^2}
{dk^{\prime 2}\over k^{\prime 2}}\left[ {11\over 2} -{N_F\over 3}
 + 6 ln(1-x^{\prime})\right]f_{g}
(x^{\prime},k^{\prime 2})\nonumber\\
&&\hspace*{10ex}{\bf(\hspace*{3ex}Altarelli - Parisi \hspace*{4ex})}\nonumber\\
&-&{\alpha_S(k^2)\over 2\pi}
\int_{x^{\prime}}^1 {dz\over z}
\Biggl(
\Biggl[ \frac{\tilde  {\bf F}_8 }{\omega^2} \Biggr](z)
\frac{ {\bf G}_0 }{2\pi^2}
\Biggr)_{gq}
\int_{k_0^2}^{k^2}
{dk^{\prime 2}\over k^{\prime 2}}
f_{S}\left({x^{\prime}\over z},k^{\prime 2}\right)\label{unifglue}\\
&-&{\alpha_S(k^2)\over 2\pi}
\int_{x^{\prime}}^1 {dz\over z}
\int_{k^2}^{k^2/z}
{dk^{\prime 2}\over k^{\prime 2}}
\Biggl(
\Biggl[\frac{\tilde   {\bf F}_8 }{\omega^2} \Biggr]
\Biggl(\frac{k^{\prime 2}}{k^2}z \Biggr)\frac{ {\bf G}_0 }{2\pi^2}
\Biggr)_{gg}
f_{g}\left({x^{\prime}\over z},k^{\prime 2}\right).\nonumber\\
&&\hspace*{10ex}{\bf(Non-ladder)}\nonumber
\eqnx
}
In equations (\ref{unifns}), (\ref{unifsea}), (\ref{unifglue})
we group separately: terms corresponding to the ladder
diagram contributions to the double $ln^2(1/x)$ resummation,
contributions from the non-singular parts of the AP
splitting functions, and finally contributions from the non-ladder
bremsstrahlung diagrams.  We label those three contributions as
"ladder", "Altarelli - Parisi " and "non-ladder" respectively.

Inhomogeneous terms $f_i^{(0)}(x^{\prime},k^2)$ ($i=NS,S, g$),
as stated above, may be expressed as~:
\eqn
f_{NS}^{(0)}(x^{\prime},k^2)&=&
{\alpha_S(k^2)\over 2 \pi}{4\over 3}
\int _{x^{\prime}}^1
{dz\over z} {(1+z^2)g_1^{NS(0)}({x^{\prime}\over z})-
2zg_1^{NS(0)}(x^{\prime})\over 1-z}\nonumber\\
&&+{\alpha_S(k^2)\over 2 \pi}\left[2 +{8\over 3}
ln(1-x^{\prime})\right]g_1^{NS(0)}(x^{\prime})
\label{fns0},
\eqnx
\eqn
f_{S}^{(0)}(x^{\prime},k^2) &=&
+{\alpha_S(k^2)\over 2 \pi}{4\over 3}
\int _{x^{\prime}}^1
{dz\over z} {(1+z^2)g_1^{S(0)}({x^{\prime}\over z})-
2zg_1^{S(0)}(x^{\prime})\over 1-z}\nonumber\\
&&+{\alpha_S(k^2)\over 2 \pi}(2 +{8\over 3} ln(1-x^{\prime}))
g_1^{S(0)}(x^{\prime})\nonumber\\
&&+{\alpha_S(k^2)\over 2 \pi}N_F\int _{x^{\prime}}^1{dz\over z}
(1-2z)\Delta g^{(0)}({x^{\prime}\over z}),\nonumber\\
f_g^{(0)}(x^{\prime},k^2) &=&
+{\alpha_S(k^2)\over 2 \pi}{4\over 3}
\int _{x^{\prime}}^1{dz\over z}(2-z)g_1^{S(0)}({x^{\prime}\over
z})\nonumber\\
&&+{\alpha_S(k^2)\over 2 \pi}({11\over 2} -{N_F\over 3}
 + 6 ln(1-x^{\prime}))\Delta g^{(0)}(x^{\prime})\label{fsg0}\\
&&+{\alpha_S(k^2)\over 2 \pi}6
\int_{x^{\prime}}^1 {dz\over z}\left[
{\Delta g^{(0)}
({x^{\prime}\over z})- z\Delta g^{(0)}
(x^{\prime})\over 1-z} +(1-2z)\Delta g^{(0)}
({x^{\prime}\over z})\right].\nonumber
\eqnx

Equations (\ref{unifns}), (\ref{unifsea}), (\ref{unifglue}) together with
(\ref{fns0}), (\ref{fsg0}) and (\ref{dpi}) reduce to the LO
AP evolution equations for
nucleon structure function with starting (integrated) distributions
$g_1^{i,(0)}(x)$ ($i=NS,S$) and $\Delta g^{(0)}(x)$ after we set the upper
integration limit over $k^{\prime 2}$ equal to $k^2$ in all terms in
equations (\ref{unifns}), (\ref{unifsea}), (\ref{unifglue}),
neglect the higher order terms in the kernels,
and set $Q^2$ in place of $W^2$ as the upper integration limit of the integral
in eq.\ (\ref{dpi}).

\section*{Appendix B}

We prove that (\ref{nsmom00}) holds. After integrating both sides of
(\ref{fns0}) over $x$ in the interval $x=(0,1)$ one arrives at~:
\eqn
{\ove f_{NS}^{(0)}(k^2)}&=&\frac{\alpha_S(k^2)}{2\pi}\,\frac{4}{3}
\int_0^1 \,dx \,g_1^{NS,(0)}(x)\,
\left[\int_0^x\,dz\,{{1+z^2}\over {1-z}}+\int_x^1\,dz\,\left({{1+z^2}\over
{1-z}}-{2 \over{1-z}}\right)\right]\nonumber\\
&+&2\,\frac{\alpha_S(k^2)}{2\pi}{\ove g_{NS}^{(0)}}+
\frac{\alpha_S(k^2)}{2\pi}\,\frac{8}{3}\int_0^1\,dx\,\ln(1-x)\,g_1^{NS,(0)}(x).
\label{ap1}
\eqnx
Performing the integrals over $z$, one obtains~:
\eqn
{\ove f_{NS}^{(0)}(k^2)}&=&\frac{\alpha_S(k^2)}{2\pi}\,\frac{4}{3}
\left[-2\int_0^1\,dx\,\ln(1-x)\,g_1^{NS,(0)}(x)-\frac{3}{2}\,
{\ove g_{NS}^{(0)}}\right]\nonumber\\
&+&2\,\frac{\alpha_S(k^2)}{2\pi}{\ove g_{NS}^{(0)}}+
\frac{\alpha_S(k^2)}{2\pi}\,\frac{8}{3}\int_0^1\,dx\,\ln(1-x)\,g_1^{NS,(0)}(x).
\label{ap2}
\eqnx
This implies~:
\eq
{\ove f_{NS}^{(0)}(k^2)}=0.
\label{ap3}
\eqx
%


\section*{Appendix C}

We prove that (\ref{fsmom0}), (\ref{fgmom0}) hold.
After integrating both sides of  (\ref{fsg0})
over $x=(0,1)$ one arrives at~:
\eqn
{\ove f_{S}^{(0)}(k^2)}&=&\frac{\alpha_S(k^2)}{2\pi}\,\frac{4}{3}
\int_0^1 \,dx \,g_1^{S,(0)}(x)\,
\left[\int_0^x\,dz\,{{1+z^2}\over {1-z}}+\int_x^1\,dz\,\left({{1+z^2}\over
{1-z}}-{2 \over{1-z}}\right)\right]\nonumber\\
&+&2\,\frac{\alpha_S(k^2)}{2\pi}{\ove g_{S}^{(0)}}+
\frac{\alpha_S(k^2)}{2\pi}\,\frac{8}{3}\int_0^1\,dx\,\ln(1-x)\,g_1^{S,(0)}(x)
\nonumber\\
&+&\frac{\alpha_S(k^2)}{2\pi}\,N_F\,\int_0^1\,{{dx}\over{x^2}}
\,\Delta p_g^{(0)}(x)\,\int_0^x\,dz\,(x-2z).
\label{bap1}
\eqnx
Using (\ref{ap2}) and performing integral over $z$ in the last term, one
obtains~:
\eq
{\ove f_{S}^{(0)}(k^2)}=0.
\label{bap2}
\eqx
In the gluon sector integration of (\ref{fsg0}) over $x$ yields~:
\eqn
{\ove f_{g}^{(0)}(k^2)}&=&
\frac{\alpha_S(k^2)}{2\pi}\,\frac{4}{3}\int_0^1\,{{dx}\over{x^2}}
\,g_1^{S,(0)}(x)\,\int_0^x\,dz\,(2x-z)\nonumber\\
&+&\frac{\alpha_S(k^2)}{2\pi}\,6\,
\int_0^1 \,dx \,\Delta p_g^{(0)}(x)\,
\left[\int_0^1\,dz\,\left({1\over {1-z}}+1-2z\right)-\int_x^1\,dz\,
{1\over{1-z}}\right]\nonumber\\
&+&\,\frac{\alpha_S(k^2)}{2\pi}\,\left(\frac{11}{2}-\frac{N_F}{3}\right)
{\ove \Delta p_g^{(0)}}+
\frac{\alpha_S(k^2)}{2\pi}\,6\,\int_0^1\,dx\,\ln(1-x)\,\Delta p_g^{(0)}(x)
\label{bap3}.
\eqnx
Performing the integrals over $z$ in (\ref{bap3}), one obtains~:
\eq
{\ove f_{g}^{(0)}(k^2)}=\frac{\alpha_S(k^2)}{2\pi}\left[{\ove g_1^{S,(0)}}
+\left(\frac{11}{2}-\frac{N_F}{3}\right){\ove \Delta p_g^{(0)}}\right]
\label{bap4}
\eqx
%


\section*{Acknowledgments}

I thank J. Kwieci\'nski for reading the manuscript and discussions 
and B. Bade\l{}ek for discussions.
This research was supported in part by the Polish Committee for Scientific
Research with grants 2 P03B 04718, 2 P03B 05119, 2PO3B 14420 and European
Community grant 'Training and Mobility of Researchers', Network 'Quantum
Chromodynamics and the Deep Structure of Elementary Particles'
FMRX-CT98-0194. B.\ Z.\ was supported by the Wenner-Gren Foundations.


\end{document}